\documentclass[onecollarge]{svjour2}
\usepackage{epsfig}
\usepackage{amsmath}
\usepackage{amssymb}
\newcommand{\bm}[1]{ \mbox{\boldmath $#1$}  }

\journalname{Few-Body Syst}

\begin{document}

\title{Dimers, Effective Interactions, and Pauli Blocking Effects in a Bilayer
of Cold Fermionic Polar Molecules}

\author{N.~T. Zinner \and J.~R. Armstrong \and A.~G. Volosniev \and D.~V. Fedorov \and A.~S. Jensen}

\institute{ Department of Physics and Astronomy,
         Aarhus University, DK-8000 Aarhus C, Denmark }

\date{\today}

\maketitle

\begin{abstract}
We consider a bilayer setup with two parallel planes of 
cold fermionic polar molecules when the dipole moments are oriented perpendicular
to the planes. The binding energy of two-body states with one 
polar molecule in each layer is determined and compared to various analytic 
approximation schemes in both coordinate- and momentum-space. The effective interaction 
of two bound dimers is obtained by integrating out the internal dimer bound state wave function 
and its robustness under analytical approximations is studied. Furthermore, 
we consider the effect of the background of other fermions on the dimer
state through Pauli blocking, and discuss implications for the zero-temperature
many-body phase diagram of this experimentally realizable system.
\end{abstract}

\section{Introduction}
The cooling and trapping of ultracold polar molecules is a rapidly growing 
field at the moment \cite{ospelkaus2008,ni2008,deiglmayr2008,lang2008,carr2009,ni2010,ospelkaus2010}, 
which generates huge theoretical interest \cite{baranov2008,lahaye2009}. 
To avoid heavy losses from chemical reactions \cite{ospelkaus2010} or 
many-body collapse \cite{lushnikov2002}, trapping in 
low-dimensional geometries has been explored in multiple works
\cite{goral2002,demille2002,barnett2006,micheli2006,wang2006,micheli2007,buchler2007,gorshkov2008,ticknor2009,micheli2010}.
Very recently, a multilayer system of 
fermionic polar molecules was experimentally realized and
chemical reactions rates that depend on the geometry and population of optical 
lattice states have been measured \cite{miranda2011}.
In addition to the potential for ultracold controlled chemistry of 
such a system, the long-range character of the dipole-dipole interaction means that the 
interactions of the molecules can potentially support a number of 
exotic few- and many-body phases
\cite{wang2007,wang2008,bruun2008,sarma2009,cooper2009,deu2010,santos2010,sun2010,yamaguchi2010,pikovski2010,zinner2010,cremon2010,baranov2011,zinner2011,levinsen2011,ticknor2011,wunsch2011,zinner2011b,artem2011c,huang2011}.

In the case of dipoles aligned perpendicular to the layers by an 
externally applied field the interaction of dipoles in different layers 
has a very interesting structure with short-range attraction and long-range
repulsion. In fact, in momentum space the interaction becomes negative 
definite in the two-dimensional (2D) limit where the layers are assumed infinitely thin
\cite{wang2008}. This suggests that the system could exhibit pairing. However, 
the intralayer interaction for the same alignment is purely repulsive \cite{fischer2006},
and this counteracts the pairing mechanism. This interplay of repulsive 
and attractive interactions in parallel planes is reminiscent of features 
in high-temperature superconductors and layers of cold polar molecules are 
therefore a very interesting model system. 

Here we focus on the case of a bilayer which already contains many of the features expected
of the general many-layer system. The two-dimensional two-component Fermi gas 
with attractive short-range interactions is a well-studied 
problem. In particular, the interesting physics of the BCS-BEC crossover from 
weakly paired atoms to strongly bound dimers was obtained some years ago \cite{miyake1983,randeria1989,randeria1990}.
It was shown that the crossover takes place when the Fermi energy of the system becomes
comparable to the binding energy of the two-body bound state between the two 
components (typically electron spins) of the generic Fermi gas. The existence of the latter state is 
guaranteed by the famous Landau criterion which states that a potential with a 
negative definite volume integral always supports a bound state \cite{landau1977}.
A bilayer with fermionic polar molecules is
similar to the 2D Fermi gas when mapping the spin to a layer index. 
The important role of the two-body bound state in 
the many-body physics of the bilayer is now clear.

The bilayer setup with perpendicular dipole moments has a two-body bound state 
which is supported by the interlayer potential whose inner part is the only 
source of attraction. However, the spatial integral of the potential 
turn out to be zero and Landau's criterion cannot be applied to secure a 
bound state at any coupling strength of the dipolar molecules. An early 
proof of the existence of a bound state even for vanishing spatial integral
was given by Simon \cite{simon1976}. In the initial stages of the theoretical
work on dipolar systems in 2D, this result appears to have been forgotten and
Gaussian approximations that predict a critical coupling strength for a
bound two-body state to arise were used \cite{yudson1997,wang2006,wang2007,santos2010}.
The energy of the bound state for the interlayer potential of a bilayer
with perpendicular dipoles was first calculated within a scattering approach in
\cite{shih2009} and by solving the Schr{\"o}dinger equation in \cite{jeremy2010,fedorov2011}. 
The highly non-trivial dependence of the bound state energy on the coupling strength
was subsequently calculated for perpendicular \cite{klawunn2010a,artem2011b,baranov2011},
and for arbitrary orientation of the dipoles with respect to the planes \cite{artem2011a}.

In this paper we consider the properties of the two-body bound dimer for dipolar
fermions in a bilayer with dipoles oriented perpendicular to the layers. The 
exact binding is obtained by numerical solution of the Schr{\"o}dinger equation
and different approximation schemes based on Gaussian and exponential wave functions
are tested against the exact result. The momentum-space wave function is then 
calculated and used to discuss an effective dimer-dimer interaction for use in 
the corresponding many-body problem. Here we find that appropriate Gaussian 
wave functions provide good analytic approximations to the exact result everywhere
except the weak-coupling limit where the wave function becomes very delocalized in 
space. Lastly, we consider the bound state in the context of the BCS-BEC crossover.
We include the Pauli blocking effect of the fermionic background by 
solving the momentum-space Schr{\"o}dinger equation and compare to a Gaussian 
approximation scheme. The binding energy of the dimer is found to decrease 
rapidly with increasing size of the background Fermi sea. The latter means that 
the (quasi)-BEC regime where the system effectively consists of bound dimers 
occupies only a small region of the phase diagram at very low density,
making it hard to reach experimentally.

\section{Model}
We consider the case of a bilayer geometry with a
single polar molecule of mass $M$ and dipole moment $\bm d$ in each layer. We assume the 
molecules are fermions, but this will only play a role later when considering the 
effect of Pauli blocking on the binding energy.
We assume that 
the dipole moments have been aligned by an external field so that they are oriented
perpendicular to the layers in which the molecules move. In this setup the 
dipolar potential is
\begin{equation}\label{pot}
V(r)=D^2\frac{r^2-d^2}{(r^2+d^2)^{5/2}},
\end{equation}
where $d$ is the interlayer distance and $D^2=\vec{\bm d}^2/4\pi\epsilon_0$ (assuming electric dipole
moments, $\vec{\bm d}$). The three-dimensional distance between the molecules is $r^2+d^2$, i.e. $r$ is the 
relative distance in the plane. For $r/d\ll 1$ the potential behaves as an oscillator, whereas
for $r/d\gg 1$ it has a repulsive $1/r^3$ tail.
Note the interesting feature that $\int d^2\bm r V(\bm r)=0$, 
which implies a highly non-trivial behavior of the binding energy for small dipole moments
\cite{simon1976,jeremy2010,klawunn2010a,fedorov2011,baranov2011,artem2011a,artem2011b}.

\begin{figure}[htb!]
\centering
  \epsfig{file=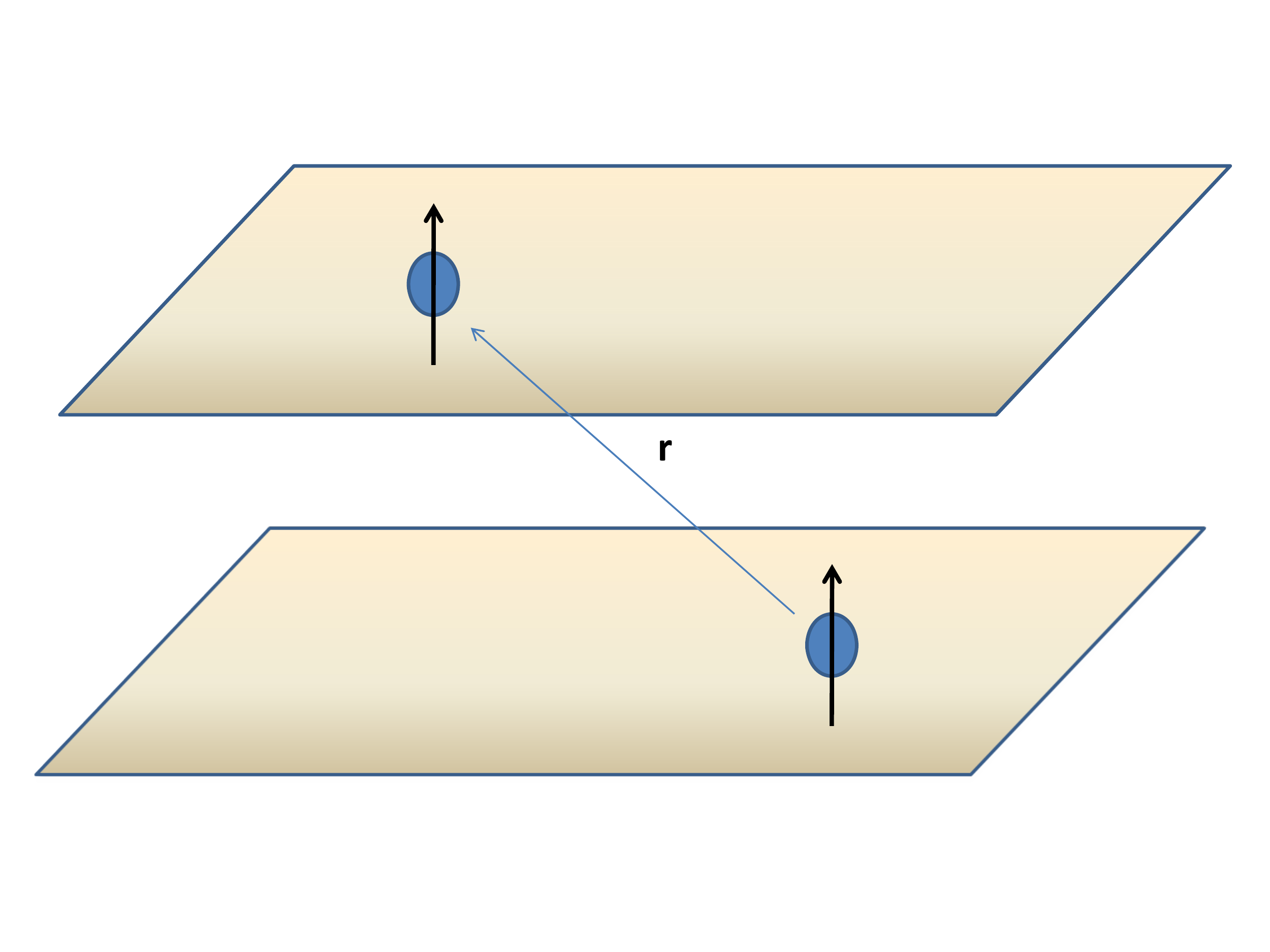,clip=true,scale=0.33}
  \caption{(color online) Schematic picture of the bilayer setup with one dipolar molecule in each 
  layer at distance $\bm r$. The dipole moment is polarized perpendicular to the bilayer planes by 
  an externally applied field.}
  \label{fig0}
\end{figure}

Since the potential has cylindrical symmetry, the two-body Schr{\"o}dinger equation can 
be decomposed into partial waves $\Psi(r)=\sum_m R_m(r)\psi_m(\phi)$ with radial equation
\begin{equation}\label{seq}
-\frac{\hbar^2}{2\mu}\left(\frac{d^2}{dr^2}+\frac{1-4m^2}{4r^2}\right)u_m(r)+V(r)u_m(r)=Eu_m(r),
\end{equation}
where $\mu=M/2$ is the reduced mass and $u_m(r)=\sqrt(r)R_m(r)$ is the reduced radial wave function. 
In this paper we will only be concerned
with the lowest $m=0$ state which yields the radial equation
\begin{equation}\label{radial}
\left[\frac{d^2}{dx^2}+\frac{1}{4x^2}-U\frac{x^2-2}{(x^2+1)^{5/2}}+\tilde E\right]u_0(x)=0,
\end{equation}
where we have rescaled the equation to dimensionless units, i.e. $x=r/d$ and $\tilde E=Md^2E/\hbar^2$.
The dimensionless strength of the dipolar interaction in the bilayer setup is $U=MD^2/\hbar^2 d$. 
Unless otherwise stated, we will use $\hbar^2/Md^2$ as the unit of energy and $d$ for lengths.

The two-body Schr{\"o}dinger equation in Eq.~\eqref{radial} can be solved numerically by standard methods. We denote
the exact energy obtained from solving this equation by
$E_0$ and the wave function by $\Psi_0$. 
However, when the dimer appears as input for more involved calculations on the many-body physics
of layered systems, it is extremely useful to have accurate analytic or semi-analytic
approximations to the full wave function. This will be our main concern in this paper, and 
we now outline four approximation schemes that we want to compare. The first three are based
on an increasingly more sophisticated oscillator approximation, whereas the last one
is an exponential ansatz for the wave function. What will be evident is that it is very 
important for any approximation to properly reproduce the spatial profile of the 
wave function.

\subsection{Scheme 1}
The simplest and also crudest approximation is to expand the potential to quadratic 
order for $r\ll d$,
\begin{equation}
V(r)\approx -\frac{2D^2}{d^3}+\frac{1}{2}\frac{12D^2}{d^5}r^2,
\end{equation}
which yields the relation $\mu \omega^2=\tfrac{12D^2}{d^5}$ for the angular frequency of 
the oscillator. The corresponding wave function is simply
\begin{equation}
\Psi_1(r)=\frac{1}{\sqrt{\pi}b}e^{-\tfrac{r^2}{2b^2}},
\end{equation}
where $b^2=\hbar/\mu\omega$.
As we will see below, this gives a bad estimate for the binding energy, but does exhibit 
some of the right behavior for large $U$ where the wave function localizes and becomes increasingly
Gaussian. We denote the energy and wave function 
obtained from this first approximation scheme by $E_1$ and 
$\Psi_1$. Recall that for Gaussian wave functions with parameter $b$ in two dimensions 
we have $\langle r^2\rangle=b^2$.

\subsection{Scheme 2}
Another strategy is to use the nice Gaussian wave function but employ the oscillator length corresponding to the 
correct binding energy. The relation between binding energy, $E_B$, and oscillator
length parameter, $l$, is $l^2=2\hbar^2/ME_B$. We now take $E_B=|E_0|$, i.e. we 
use the {\it exact} result for the energy to determine the length scale of the 
Gaussian ground-state wave function $\Psi_2$. This has the advantage that the wave function 
will become increasingly extended for small $U$ and, in turn, small $|E_0|$.

\subsection{Scheme 3}
The third use of Gaussian approximation schemes is somewhat more involved and builds
on ideas discussed earlier in Ref.~\cite{jeremy2010}. We approximate the true potential 
by a model potential which is quadratic but with a constant (negative) shift. We then 
fit the oscillator frequency and the shift to ensure that (i) the binding energy is 
equal to the exact result and (ii) the node of the potential occurs at $r/d=\sqrt{2}$
which is the same position as for the real potential in Eq.~\eqref{pot}. The latter
condition reflects the idea that the inner attractive part is the important one 
that determines the properties of the system, whereas the outer repulsive tail of the
true dipolar interaction has only limited influence. Assuming a potential of 
the form $V(r)=M\tilde\omega^2r^2/4-E_s$, the two conditions yield the 
following matching relations between angular frequency, $\tilde \omega$, shift, $E_s>0$, 
and $E_0$
\begin{equation}
\frac{1}{2}Md^2\tilde\omega^2 = E_s \quad \text{and}\quad E_0=\hbar\tilde \omega-E_s.
\end{equation}
These second order equations have two solutions 
\begin{equation}
x_\pm=1\pm\sqrt{1-2\tilde E_0}, 
\end{equation}
where $x_\pm=Md^2\tilde\omega_\pm/\hbar$ and $\tilde E_0=Md^2E_0/\hbar^2$.
The
solutions have the properties that $x_+\to 2-E_0$ and $x_-\to E_0$ for $|E_0|\to 0$.
We see that the $x_+$ solution goes to a constant oscillator frequency, whereas the 
$x_-$ goes to a vanishing frequency. 
The approximating potential becomes
\begin{equation}
V(r)=\left(x_\pm-\frac{Md^2E_0}{\hbar^2}\right)\left[\frac{1}{2}\left(\frac{r}{d}\right)^2-1\right],
\end{equation}
where since $E_0<0$ the first factor is always positive and tends to zero for $|E_0|\to 0$
in the weak-coupling limit. The solution $x_-$ is negative, and 
therefore $\tilde\omega_-=\hbar x_-/Md^2<0$. Therefore this quantity cannot be interpreted as an 
oscillator frequency as it stands. However, if we take $\tilde\omega_-=\hbar|x_-|/Md^2$, 
we get a positive frequency which has the desirable feature that the wave function 
is allowed to extend in space when $|E_0|$ is small. The price we pay is that 
now the binding energy in the oscillator is not equal to the exact result, $E_0$, but
becomes
\begin{align}
Md^2\tilde\omega_-/\hbar-Md^2E_s/\hbar^2=2\sqrt{1-2\tilde E_0}+\tilde E_0-2,
\end{align}
in dimensionless form. The energy is the $\tilde\omega_-$ oscillator is then positive
for $-\tilde E_0<4$, but still goes to zero as $|E_0|\to 0$. Since it is highly desirable
to have an extended potential (i.e. $\tilde\omega_-\to 0$) as $|E_0|\to 0$ we retain
this solution despite the difference in ground state energy.
The length scale of the Gaussian ground-state of the $\tilde\omega_\pm$ oscillators has the simple form $l^2/d^2=2/|x_\pm|$, 
which clearly diverges as $|E_0|\to 0$ for the $\tilde\omega_-$ solution. 
We denote the corresponding wave function $\Psi_3$.

Below we will see that whereas the approximation using $\tilde\omega_-$ compares poorly
with exact results, using $\tilde\omega_+$ gives a much better approximation in spite of 
its localized behaviour when approaching zero energy ($x_+\to 2$). The approximation 
with $\tilde\omega_+$ has been used to calculate the bound state energy of chains with 
one molecule in each layer in the three- and four-layer cases, and yields results
that are in excellent agreement with exact numerical methods \cite{armstrong2011,artem2012,armstrong2012}.

\subsection{Scheme 4}
The final approximation scheme uses an exponential wave function instead of a Gaussian but
still uses the exact binding energy. This means that we reproduce the correct 
large-distance behavior of the exact wave function. The wavenumber $\kappa$ of the exponential is fixed similar
to approximation scheme 2, i.e. $|E_0|=\hbar^2\kappa^2/2\mu$ or $\kappa=\sqrt{M |E_0|/\hbar^2}$.
The properly normalized wave function for this scheme is
\begin{equation}
\Psi_4(r)=\sqrt{\frac{2}{\pi}}\kappa e^{-\kappa r}.
\end{equation}
For this wave function we have $\langle r^2\rangle=3/2\kappa^2$. 
Scheme 4 is special in the sense that it is designed to approximate the 
tail behavior of the full wave function. Since the exponential form is not correct at small distances it does
not make sense to calculate the energy for this wave function. The correct wave function for bound states
outside the range of the potential in two dimensions is the modified Bessel function of the second 
kind $K_0$, which does have the exponential tail but is logarithmically divergent at the origin.

\section{Dimer Properties}
We now compare the properties of the exact dimer solution from the Schr{\"o}dinger equation to the 
various approximation schemes introduced above. We do this for the energy obtained in scheme 1, and for the
wave function and radius squared for all schemes. Furthermore, we consider the Fourier transform of 
the wave functions and of their square which will be of use when discussing the effective interaction 
of dimers below.

\subsection{Binding Energy}
\begin{figure}[htb!]
\centering
  \epsfig{file=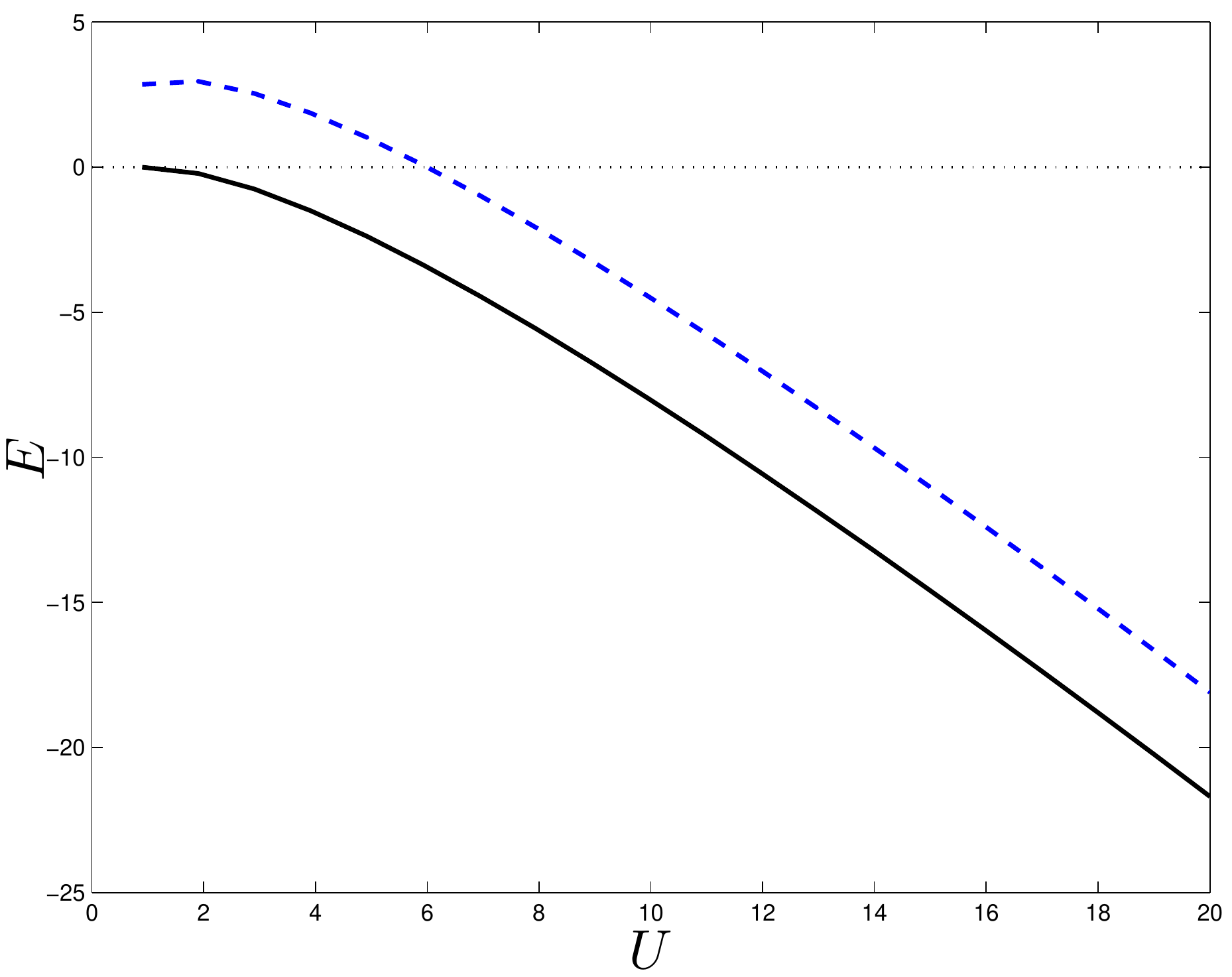,clip=true,scale=0.45}
  \caption{(color online) Dimer energy as a function of $U$ in units of $Md^2/\hbar^2$ for $U\geq 0.9$. The solid (black) line is the 
  exact solution ($E_0$), whereas the dashed (blue) line is the energy obtained by expanding the dipolar potential around the origin
  to second order in approximation scheme 1 ($E_1$). The horizontal dotted (black) line indicates zero energy.}
  \label{fig1a}
\end{figure}

The first property of the dimer system to study is the binding energy. Since the ground-state has cylindrical symmetry, 
the exact solution can be obtained by integrating the Schr{\"o}dinger equation and matching to a bound state
wave function at large distances. This energy has been calculated by several groups in recent 
years \cite{shih2009,jeremy2010,klawunn2010a,baranov2011,artem2011a}. In Fig.~\ref{fig1a} the energy, $E_0$, is plotted
as a function of the dipolar strength $U$. At very small $U\sim 1$ one sees the rapid decrease of the binding
energy which to leading order goes as $e^{-8/U^2}$ \cite{simon1976,klawunn2010a,baranov2011,artem2011a,artem2011b}.
The calculation was done for $U\geq 0.9$ as smaller $U$ values have binding energies that were below the numerical 
precision. Note that the exact energy is recovered by definition in approximation
schemes 2 and 3 above ($E_2=E_3=E_0$). 

Also shown in Fig.~\ref{fig1a} is the result of calculating the energy by a crude harmonic approximation to the 
potential around $r=0$. The energy has the following analytical expression
\begin{equation}
\frac{Md^2}{\hbar^2}E_1 = \sqrt{24U}-2U.
\end{equation}
For $U<6$ we find $E_1>0$. This is of course due to the zero-point energy which is very large in this particular
approximation. However, at larger $U$ we do see the correct slope of $E_1$ as compared to the exact result, and 
the difference seems to be only an overall offset. This clearly indicates that approximating the wave function 
by a Gaussian should be an accurate description for larger $U$ as noted in \cite{jeremy2010}. It is of course
clear that this tight harmonic approximation will fail at small $U$. To remedy this situation, we now 
discuss a number of alternative approximations to the wave function which use knowledge of the exact result,
but where the scale of the wave function varies with $U$ in a different way as compared to the crude harmonic 
approximation of $\Psi_1$.

First, we check the energies calculated within the various approximation that are based on Gaussians where a nice analytical expression 
can be given for the total energy. In fact, we have 
\begin{align}
&\frac{Md^2}{\hbar^2}\langle H \rangle = x^2 \left[1- U f(x)\right],&\label{hamil}\\
&f(x)=-4x^2+2\sqrt{\pi}\text{Erfc}(x)e^{x^2}(2x^3+x),&
\end{align}
where $\text{Erfc}(x)$ is the complementary error function. We used the quantity $x=d/b$, where $b$ is the length parameter
in the Gaussian wave function. The optimal variational energy is now found by finding the minimum of Eq.~\ref{hamil} with
respect to $x$. This value can be compared to the choices of length parameter based on the exact energy and the shape of
the potential as outlined in approximation schemes 2 and 3. We have chosen not to list this approximation among the 
schemes above since it is not based on the knowledge of the binding energy for construction of the wave function
which is what the comparison done here is about. The variational approach provides both a binding energy and a 
corresponding (Gaussian) wave function, which is as we will now see close to some of the other schemes.

\begin{figure}[htb!]
\centering
  \epsfig{file=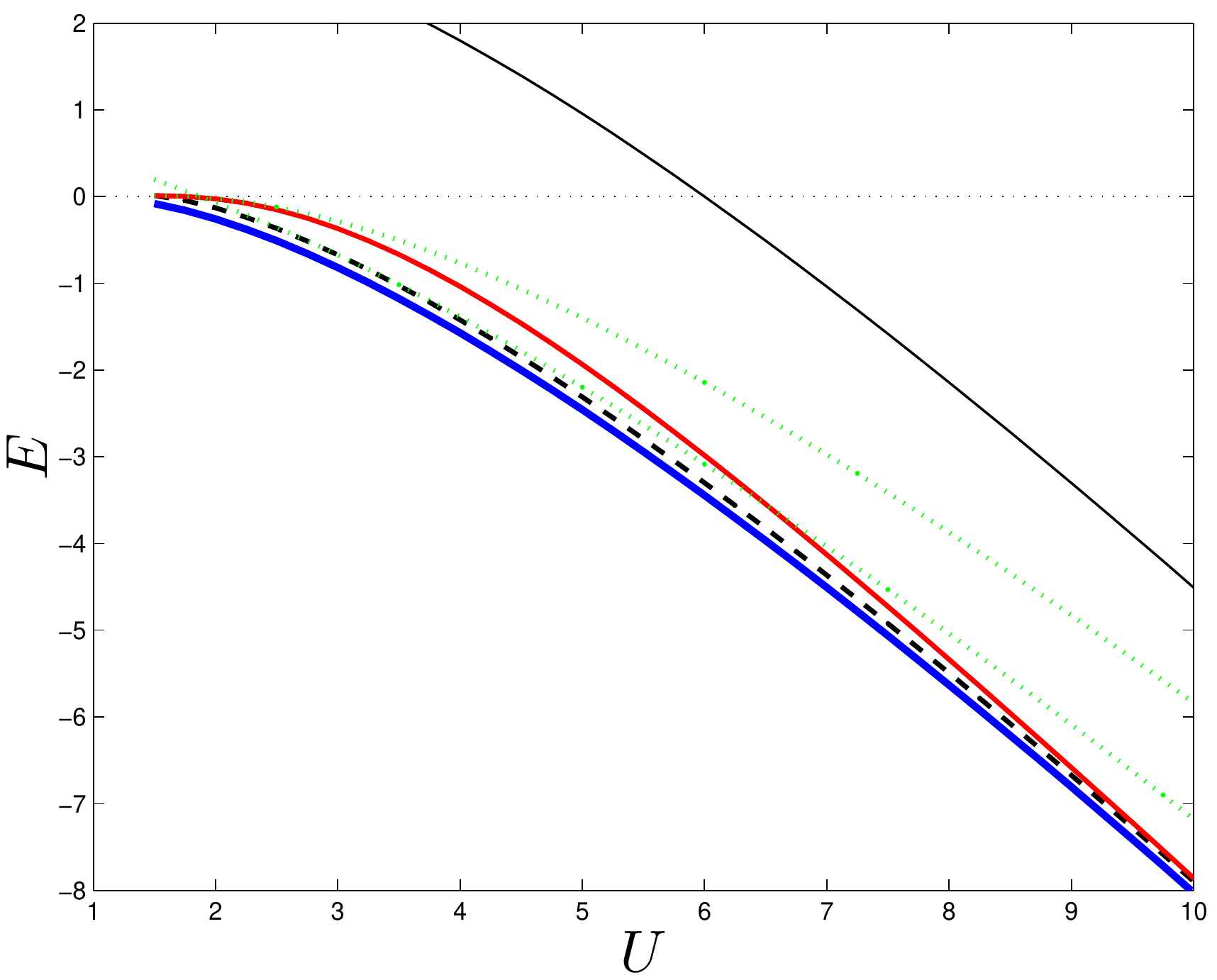,clip=true,scale=0.45}
  \caption{(color online) Dimer energy as a function of $U$ in units of $Md^2/\hbar^2$ for $U\geq 0.9$. The thick solid (blue) line is the 
  exact solution ($E_0$) and the upper thin solid (black) line is the energy obtained by expanding the dipolar potential around the origin
  to second order in approximation scheme 1 ($E_1$). The solid (red) line is the energy based on the choice of length parameter 
  in scheme 2, while the dashed (black) line is the optimum variational choice when using a Gaussian wave function. The two dotted (green) 
  lines are for scheme 3 with the lower curve at large $U$ corresponding to $\tilde\omega_+$ and the upper one to $\tilde\omega_-$.  
  The vertical dotted (black) line indicates zero energy.}
  \label{fig1b}
\end{figure}

In Fig.~\ref{fig1b} we plot the energies of the different Gaussian
approaches along with the exact energy and also the energy result of the crude approximation in scheme 1. Note that all 
approximate solutions have positive energy below $U\sim 1.5$.
Comparing the 
solid (red) and dashed (black) curves we see that the variationally optimized solution is better than scheme 2 but not by
a large margin, and the difference becomes small for large $U$ where they both approach the exact result. The two dotted
(green) curves from scheme 3 are interesting as $\tilde\omega_-$ seems to do consistently worse than $\tilde\omega_+$, with the exception of 
small $U$ where the $\tilde\omega_+$ solution goes positive. Note how scheme 2 and scheme 3 with $\tilde\omega_+$ does well
in the large $U$ and small $U$ regimes respectively, where they agree also with the variational calculation.

\subsection{Wave function and Size} 
\begin{figure*}[ht!]
\centering
  \epsfig{file=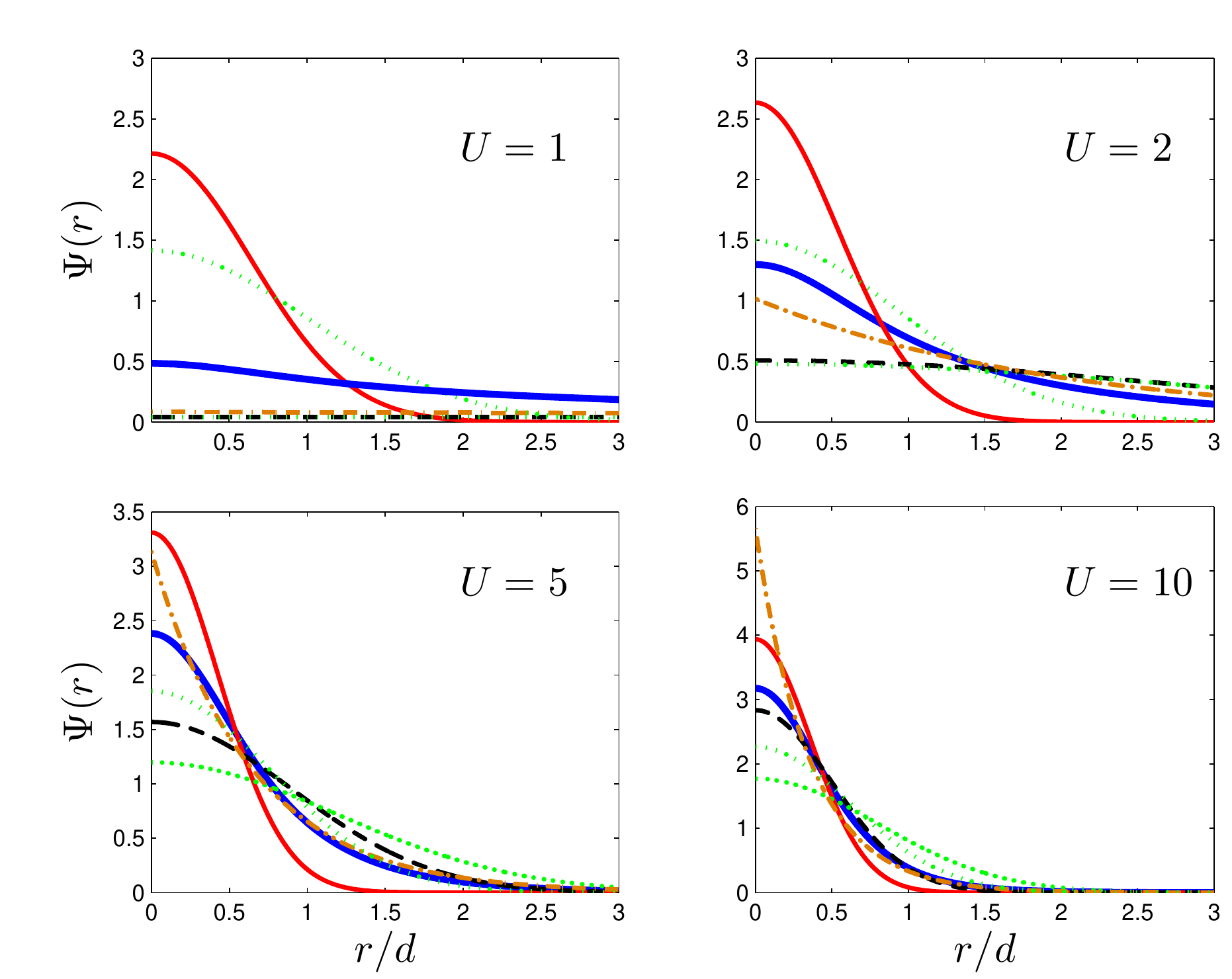,clip=true,scale=0.75}
  \caption{(color online) Dimer radial wave function for $U=1$, 2, 5 and 10. in units of $Md^2/\hbar^2$. The solid thick (blue) line is the 
  exact solution. The other lines are approximation scheme 1 (solid red), 2 (dashed black), 3 (dotted green), and 4 (dash-dotted brown).
  Note the difference in scales on the vertical axis. The upper (at $r=0$) dotted green corresponds to $\tilde\omega_+$ in scheme 3 and the lower one 
  to $\tilde\omega_-$.}
  \label{fig2}
\end{figure*}

Next we look at the wave function of the exact solution in comparison to the different approximation schemes. In Fig.~\ref{fig2}
we plot all the different choices for strengths $U=1$, 2, 5, and 10. As expected, scheme 1 is poor for small $U$ and only
becomes acceptable around $U=10$, while the exponential approximation to the wave function is quite good in all cases.
The former scheme simply can not deal with a rapid spreading of the wave function at small $U$.  
For schemes 2 and 3 we see quite good agreement with scheme 2 always giving a slightly better approximation.

\begin{figure}[htb!]
\centering
  \epsfig{file=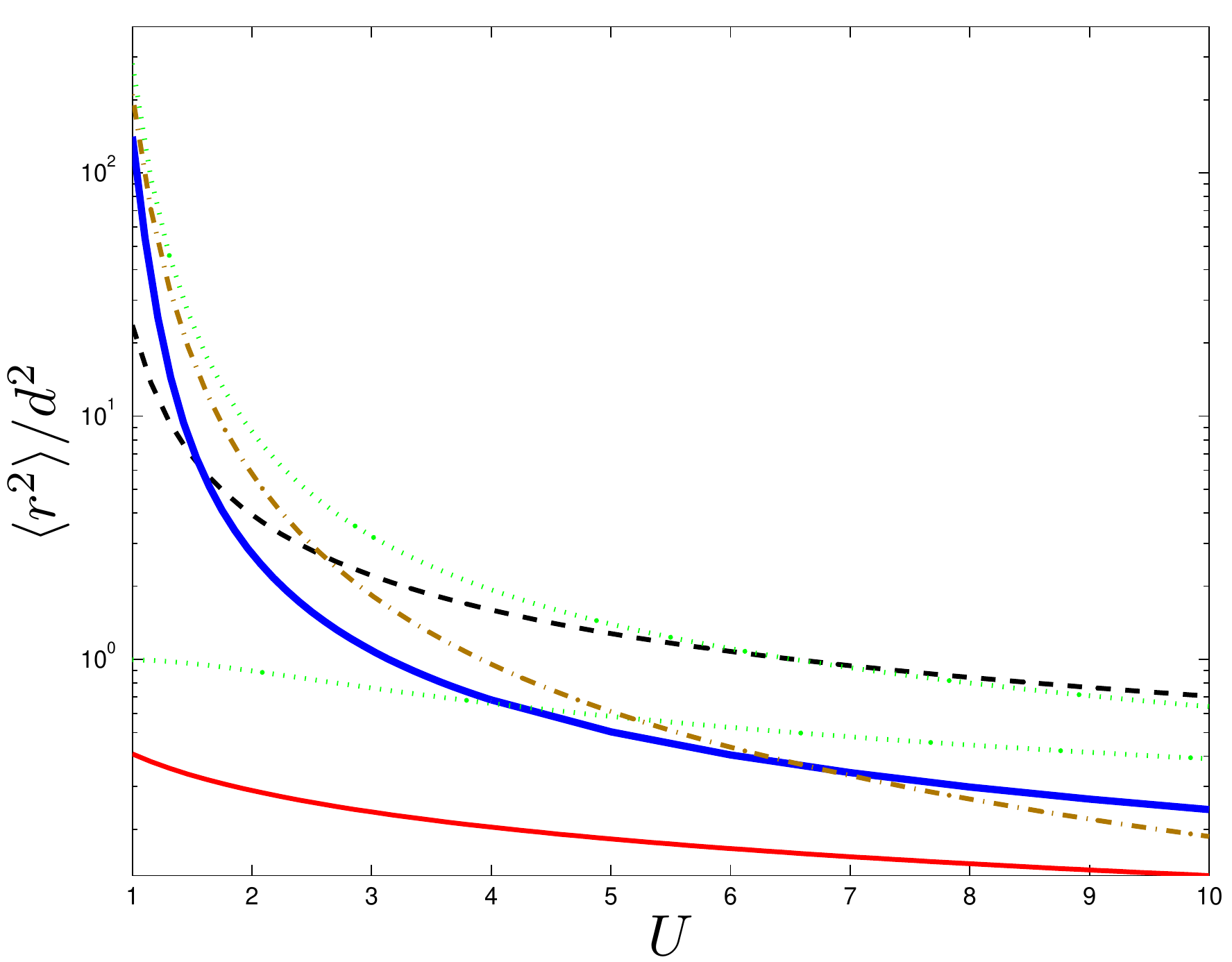,clip=true,scale=0.45}
  \caption{(color online) Mean radius squared, $\langle r^2\rangle$, as function of $U$ on a logarithmic scale. The lines 
  are as in Fig.~\ref{fig2}. The upper (at $U=1$) dotted (green) line corresponds to $\tilde\omega_-$ and the lower one to $\tilde\omega_+$ within scheme 3.}
  \label{fig3}
\end{figure}

From the wave function we can determine the extension of the dimers in the various approximation by computing $\langle r^2\rangle$. The results
are shown in Fig.~\ref{fig3} on a logarithmic scale which is convenient since at $U\sim 1$ the dimer becomes extremely extended. 
Again we see that the exponential form in scheme 4 does remarkably well, whereas scheme 2 and 3 overestimate the size for 
$U\gtrsim 2.5$. Scheme 1 has a very small radius for all $U$ and is quite far off the exact result as was clear from the wave functions
in Fig.~~\ref{fig2}. It is worth noting that all curves in Fig.~\ref{fig3} have similar decreasing behavior for large $U$ although the
asymptotic values are very different. It is also interesting that even though the exact result predicts a dimer of size less than 
the attractive part of the potential for $U\gtrsim 4$, scheme 1 does not provide a good size estimate until much larger $U$ even though
it was built on only this inner attractive harmonic part of the full dipole-dipole potential. The lowest order expansion is 
therefore clearly insufficient.

From the results above, we can conclude that the key issue to obtain a good
approximation seems to be the choice of bound-state size. Schemes 2 and 4 which includes only the 
size estimated through the binding energy are always the more accurate choices, with scheme 3 
which includes the zero of the original potential also doing well. The naive harmonic approximation 
to the potential seems to fail in all aspects as seen in Figures \ref{fig1b}, \ref{fig2}, and \ref{fig3}.

\subsection{Fourier Transforms}
Below we are concerned with the effective interaction between dimers which can be calculated
from the Fourier transform of the dimer wave function. We define the Fourier transform to momentum space as
\begin{align}
\tilde\Psi(\bm k)=\int d^2\bm x e^{-i \bm k\cdot \bm x}\Psi(\bm x).
\end{align}
Since the wave function has cylindrical symmetry we can perform the angular integration and obtain
\begin{align}
\tilde\Psi(k)=2\pi\int dx x J_0(kx) \Psi(x),
\end{align}
where $J_0(z)$ is the Bessel function of order zero. For the Gaussian wave function, $\Psi(x)=e^{-x^2/2b^2}/\sqrt{\pi b^2}$,
the result is 
\begin{align}
\tilde\Psi(k)=2\sqrt{\pi}b e^{-k^2b^2/2},
\end{align}
while the exponential,$\Psi(x)=\sqrt{2/\pi}\kappa e^{-\kappa x}$, yields
\begin{align}
\tilde\Psi(k)=2\sqrt{2\pi}\frac{\kappa^2}{(k^2+\kappa^2)^{3/2}}.
\end{align}
The remarks made about the various approximation schemes in coordinate space in relation to Fig.~\ref{fig2}
hold for the Fourier transforms as well with scheme 1 doing quite bad while scheme 2,3, and 4 comparing
much better with the exact result.

\section{Effective Dimer-Dimer Interaction}
We now turn our attention to the evaluation of an effective dimer-dimer interaction that takes both
intra- and inter-layer interactions into account and uses the full dimer wave function as well. This 
procedure was discussed in \cite{zinner2010} within the approximation scheme 2 outlined above. 
Here we elaborate and present a full comparison of various approximation and the influence 
on the effective dimer-dimer interaction.

\subsection{Derivation}
Consider two dimers in a bilayer. Let the coordinates of the polar molecules in each layer be $\bm r_1,\bm r_2$ in the first 
dimer and $\bm r_3,\bm r_4$ in the second dimer. Define the coordinates relative coordinates $\bm r=\bm r_1-\bm r_2$ and 
$\bm r'=\bm r_3-\bm r_4$, and the center-of-mass (CM) for each dimer, $\bm R=(\bm r_1+\bm r_2)/2$ and 
$\bm R'=(\bm r_3+\bm r_4)/2$. The distance between the CM of the two dimers is $\bm \rho=\bm R-\bm R'$.
The effective potential is obtained by integrating over the wave function of the dimers and over the 
CM coordinates with the condition that $\bm\rho=\bm R-\bm R'$ (we specify the meaning of this below). 
We have 
\begin{align}
&V_{eff}(\bm \rho)=\int d\bm r d\bm r' d\bm\rho'|\Psi(\bm r_1,\bm r_2,\bm r_3,\bm r_4)|^2\times &\nonumber \\
&\left[V_\text{dip}(\bm r_1-\bm r_2)
+V_\text{dip}(\bm r_3-\bm r_4)+V_\text{dip}(\bm r_1-\bm r_3) \right.&\nonumber\\
&\left.V_\text{dip}(\bm r_1-\bm r_4)+V_\text{dip}(\bm r_2-\bm r_3)+V_\text{dip}(\bm r_2-\bm r_4) \right],&
\end{align}
where $V_\text{dip}$ is the dipole potential in coordinate-space and $\Psi$ is the total wave function of the system. Here we
have defined the total center-of-mass coordinates of all four molecules, $\bm\rho'=(\bm R+\bm R')/2$. For the latter we assume
\begin{align}
\Psi(\bm r_1,\bm r_2,\bm r_3,\bm r_4)=\phi(\bm r)\phi(\bm r') \psi(\bm R)
\psi(\bm R'),
\end{align}
where $\phi$ is the relative wave function and $\psi$ is the center-of-mass wave function of the dimers. We expect this to be a good approximation for a system of many particles when the dimer is strongly bound for large $U$ and can be considered the effective constituent. We can write
\begin{align}
&V_{eff}(\bm \rho)=\int d\bm r d\bm r' d\bm\rho' |\phi(\bm r)|^2|\phi(\bm r')|^2\times &\nonumber\\
&  |\psi(\bm R)|^2 |\psi(\bm R')|^2\left[V_{\perp}(\bm r)+V_{\perp}(\bm r')+V_{\parallel}(\bm \rho+\frac{\bm r-\bm r'}{2}) \right.&\nonumber\\
&\left.V_{\parallel}(\bm \rho-\frac{\bm r-\bm r'}{2})
+V_{\perp}(\bm \rho+\frac{\bm r+\bm r'}{2})+V_{\perp}(\bm \rho-\frac{\bm r+\bm r'}{2}) \right],&
\end{align}
where $V_\perp$ is the inter-layer dipole potential and $V_\parallel$ is the intra-layer dipole potential. Assume now that 
the center-of-mass part is proportional to $\delta(\bm R-\bm R'-\bm \rho)\delta((\bm R+\bm R')/2-\bm R_\text{CM})$, where $\bm R_\text{CM}$ is the
center-of-mass position of the total four-body system which is unimportant here (we can fix the coordinate system so that $\bm R_\text{CM}=0$ in any case). One can then drop the center-of-mass part of the integral. Likewise, 
the integral over the first two potential terms gives the potential energy of the dimer which is a constant that can be discarded. We are
left with
\begin{align}
&V_\text{eff}(\bm \rho)=\int d\bm r d\bm r'  |\phi(\bm r)|^2|\phi(\bm r')|^2 \left[
V_{\parallel}(\bm \rho+\frac{\bm r-\bm r'}{2})\right.&\nonumber\\
&\left.+V_{\parallel}(\bm \rho-\frac{\bm r-\bm r'}{2})
+V_{\perp}(\bm \rho+\frac{\bm r+\bm r'}{2})+V_{\perp}(\bm \rho-\frac{\bm r+\bm r'}{2}) \right].&
\end{align}
The Fourier transform is given by
\begin{align}\label{veff}
&V_\text{eff}(\bm k)=\int V_\text{eff}(\bm \rho) \exp^{-i \bm k\cdot \bm \rho}d\bm \rho=&\nonumber\\
&\int d\bm r d\bm r'  |\phi(\bm r)|^2|\phi(\bm r')|^2
\left[2V_\parallel(\bm k)\cos\left(\bm k\cdot \frac{\bm r+\bm r'}{2}\right)\right.&\nonumber\\
&\left.+2V_\perp(\bm k)\cos\left(\bm k\cdot \frac{\bm r-\bm r'}{2}\right) \right],&
\end{align}
where $V_\parallel(\bm k)$ is the intra-layer and $V_\perp(\bm k)$ is the inter-layer momentum-space potential respectively.

\begin{figure*}[ht!]
\centering
  \epsfig{file=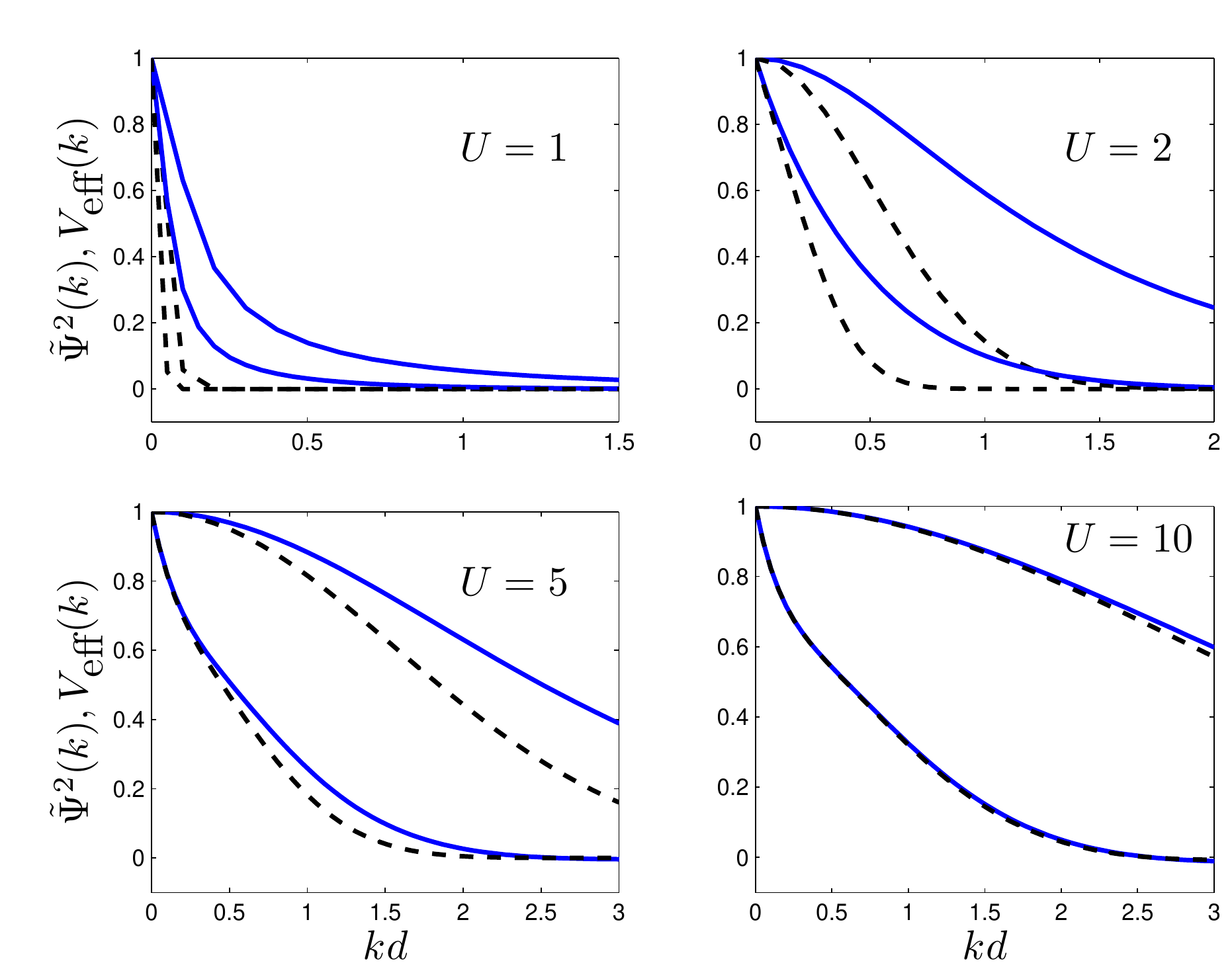,clip=true,scale=0.75}
  \caption{(color online) Plot of the (normalized) effective dimer-dimer interaction, $V_\textrm{eff}(k)/V_\textrm{eff}(0)$, along with $\tilde\Psi^2(k)$ for $U=1$, 2, 5, and 10, and with layer width $w/d=0.1$. The lower of the two solid lines is the effective potential while the upper one is $\tilde\Psi^2(k)$ (and likewise for the dashed lines). The solid (blue) lines use the exact dimer wave function, whereas the dashed (black) lines 
  are within the Gaussian approximation scheme 2. Note the differences in the horizontal scale of the panels.}
  \label{fig6}
\end{figure*}

The explicit expression for the interaction potentials in momentum space are
\begin{align}
&V_\parallel(\bm q)=\frac{8\pi D^2}{3\sqrt{2\pi}w}\left(1-\tfrac{3}{2}F(|w\bm q|)\right)&\\
&V_\perp(\bm q)= -2\pi D^2|\bm q|e^{-|\bm q|d},&
\end{align}
where $F(x)=\sqrt{\pi/2}x[1-\text{Erf}(x/\sqrt{2})]e^{x^2/2}$ with
$\text{Erf}(x)$ the error function. We have assumed that only the 
ground-state of the optical lattice potential that creates the 2D geometry
is occupied and we denote the transverse width by $w$. For the interlayer 
interaction we use the strict 2D limit result which is a very good 
approximation for the values $w\leq 0.3d$ to be used below.

\subsection{Results}
\begin{figure*}[ht!]
\centering
  \epsfig{file=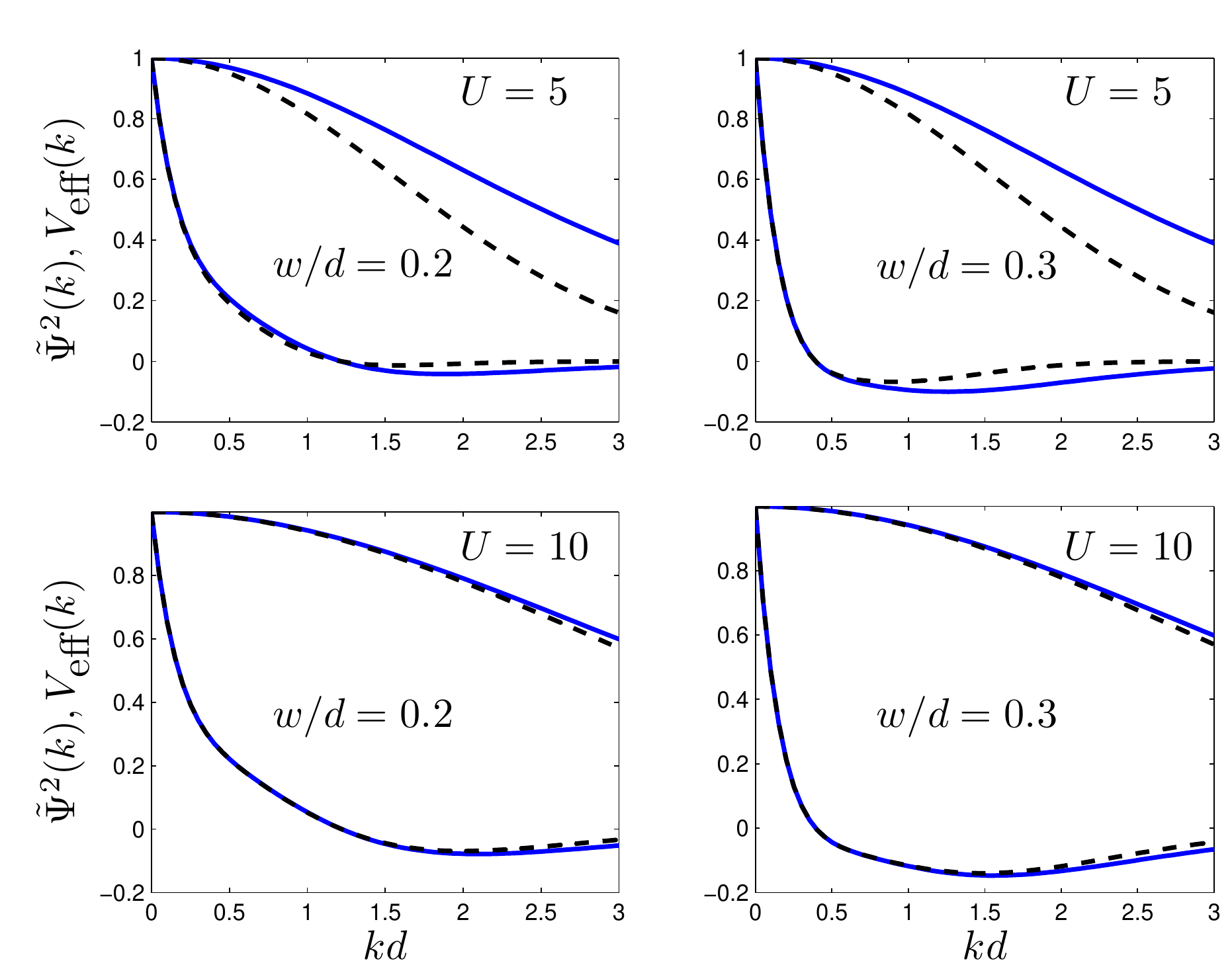,clip=true,scale=0.75}
  \caption{(color online) Same as Fig.~\ref{fig6} but with layer widths $w/d=0.2$ (left panels) and $w/d=0.3$ (right panels) for $U=5$ (upper panels) and $U=10$ (lower panels).}
  \label{fig7}
\end{figure*}

From the expression in Eq.~\eqref{veff}, it is clear that to obtain the effective interaction
of two dimers, one need the Fourier transform of the coordinate-space wave function squared. 
In Fig.~\ref{fig6}, the effective interaction based on the 
exact wave function is compared to approximation scheme 2 for a layer width of $w=0.1d$. 
As expected, the solid and dashed curves are clearly seen to approach each other as $U$ increases.
To explore the effect of the layer width, we plot $V_\text{eff}(k)$ in Fig.~\ref{fig7} 
for $U=5$ and 10 for widths $w=0.2d$ and $w=0.3d$. The agreement of the 
exact and approximate solution is striking except for $U=5$ and $w=0.3d$ where the 
Gaussian wave function seems to somewhat underestimate the attraction at large $kd$. Note
that for the physics to remain effectively 2D, we must have $kw\ll 1$ or $kd\ll d/w$. 
The range becomes smaller at larger $w$ and at large $kd$ the results are strongly modified
by the 3D dipolar physics.

We have demonstrated that the Gaussian approximation in scheme 2 is accurate at large $U$.
In particular, this was the limit in which this approximation was used 
in \cite{zinner2010} to calculate properties of the dimerized system at both zero and finite temperature.
Furthermore, it is clear that this fact holds for a range of layer sizes, and the Gaussian 
approximation can be used to study the behavior of the roton instability of a bosonic 
dimer system that should emerge in the strong-coupling large $U$ limit. The use of 
softer lattice that have larger $w$ implies more attraction at large $kd$ and the 
roton instability could become accessible at lower $U$ values.

\begin{figure}[htb!]
\centering
  \epsfig{file=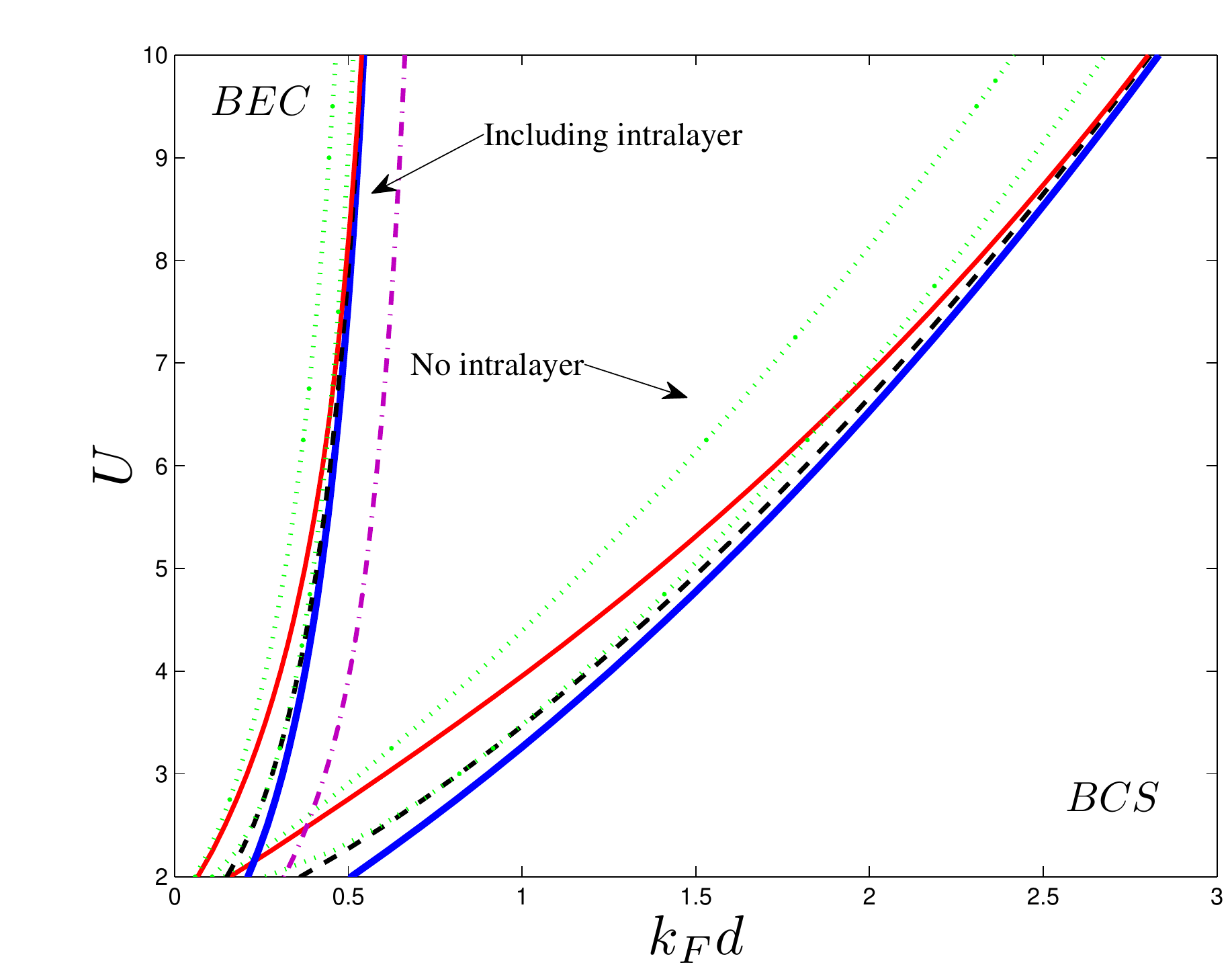,clip=true,scale=0.45}
  \caption{(color online) Zero-temperature phase diagram with lines of vanishing chemical potential both with and without including the 
  intralayer interaction as discussed in the text. Results based on the exact energy are shown as the thick solid (blue) line.
  Also shown energies based on the the variationally optimized Gaussian (dashed (black) line), scheme 2 (solid (ref) line), and scheme 3 (dotted (green) lines). For comparison, the dash-dotted (purple) line to the right of the lines marked 'Including intralayer' shows the result of a calculation of the crossover using the BCS gap equation including the self-energy term self-consistently (taken from \cite{zinner2010}). It demonstrates the good agreement between BCS (traditionally weak-coupling theory) and the strong-coupling approach discussed in the current presentation. All calculations use $w/d=0.2$.}
  \label{fig8}
\end{figure}

\section{Many-body Physics}
We now consider the many-body physics of the bilayer at intermediate to strong coupling where the BCS-BEC 
crossover is expected to happen. As discussed in the introduction, the two-component Fermi gas in 2D with 
zero-range interactions in the BCS approximation has a closed analytical solution \cite{miyake1983,randeria1989,randeria1990}
\begin{align}
&\Delta=\sqrt{2E_F E_B}&\\
&\tilde\mu=E_F-\frac{1}{2}E_B,&
\end{align}
where $\Delta$ is the constant energy gap, $\tilde\mu$ the chemical potential (recall that $\mu$ is reserved for the reduced mass), 
$E_F=\hbar^2k_{F}^{2}/2M$ the Fermi energy, and 
$E_B$ the binding energy of the two-body bound state that the attractive zero-range potential supports. We define the 
crossover from weak-coupling BCS behavior to a two-body bound state BEC as the point where $\tilde\mu=0$. If we write $E_B=\hbar^2/Ml^2$, 
with $l$ the size of the bound state, then $\tilde\mu=0$ occurs when $2\pi nl^2=1$, i.e. when the inter particle distance
is roughly the size of the bound state. In our case with a bilayer of dipolar fermions, we expect these conclusions to 
be modified by the fact that we have long-range interactions that are both attractive and repulsive.

From a Fermi liquid theory point of view, one could argue that since the long-range dipolar interaction is
present on both side of the crossover, it provides a constant shift of the chemical potential and 
a modification of the effective mass entering the two-body bound that cancel each other and have
no influence on the position where the crossover is expected. While this would be correct for 
small $U$, we are here considering the strong-coupling limit and this conclusion may no longer
be true. In fact, recent studies have found that other many-body phases in the system such 
as the density-wave state are very strongly influence by the self-energy correction which 
drives the instability to larger values of $U$ \cite{yamaguchi2010,zinner2011,babadi2011,parish2011,sieberer2011,block2011,chan2010}.
We expect qualitatively similar behaviour in relation to the pairing properties of the system. We therefore
consider the problem from the strong-coupling point of view via an effective dimer approach.

In a first approximation, we discard the intralayer interaction and consider the $\tilde\mu=E_F-\tfrac{1}{2}|E_0|=0$ 
line with $E_0$ the binding energy of the bound state caused by the interlayer potential. Lines of vanishing $\tilde\mu$ 
can be seen to the right in Fig.~\ref{fig8} for the exact result and the different approximations to the 
binding energy. Except for scheme 3, the approximations are all very close the exact result, which is not surprising
since the approximations are in one way or another derived from the demand that the size of the bound state is close
to the exact result. 

The intralayer interaction must of course be taken into account, in particular in the strong-coupling regime. One 
way to proceed is to calculate the effective mass, $m^*$ and replace $m\to m^*$ in $E_F$ as discussed above. This can be 
easily done in the weak-coupling regime where the result is analytic \cite{baranov2011,chan2010,kestner2010}.
However, we are interested in the strongly-coupled regime where we expect the dimer to be strongly bound and
hence the relevant degree of freedom. We therefore consider instead the physics of an interacting Bose gas of 
two-body bound states for which the chemical potential to lowest order in many-body perturbation theory is 
$\tilde\mu=\tfrac{1}{2}nV_\text{eff}(0)$ \cite{fetter1971}. The factor of one-half is needed since we are interested in 
the chemical potential per molecule and not per dimer. 

In order to include the intralayer interaction we consider now the chemical potential
\begin{align}\label{strong}
\tilde\mu=E_F+\frac{1}{2}n V_\text{eff}(0)-\frac{1}{2}|E_0|.
\end{align}
The lines where this expression vanishes are shown on the left side of Fig.~\ref{fig8}. Clearly the extra repulsion
at long distances (small $kd$) pushes the chemical potential up and for the binding energy to compensate this 
effect we need larger $U$, i.e. the lines are pushed to the left to low densities. Again we see that the 
exact result and the approximations for the binding energy are very close. In Fig.~\ref{fig8} we have used 
$w=0.2d$. Using a smaller $w$ will move the lines even further left and vice versa for larger $w$. Tuning 
$w$ could therefore be a necessary and convenient way to enlarge the BEC regime for experimental access.

For comparison, we also show in Fig.~\ref{fig8} a calculation of the line of vanishing chemical potential 
using BCS theory, i.e. by solving the gap equations including both inter- and intralayer interactions
as discussed in Ref.~\cite{zinner2010}. It is shown as a (purple) dash-dotted line on the right side of the 
strong-coupling approach including the intralayer term. The lines show that one finds 
agreement between the BCS approach and the strong-coupling approach discussed in the current paper. This 
suggests that the inclusion of the intralayer interaction through the $nV_\text{eff}(0)$ term in the
strong-coupling expression of Eq.~\eqref{strong} is consistent with the BCS result. Note that the 
intralayer interaction is included self-consistently in the BCS equation with the full momentum-dependent
potential \cite{zinner2010}. The intralayer interaction indeed modifies the BCS phase and not merely by 
a shift of the chemical potential proportional to $nV_\text{eff}(0)$.

The use of lowest order perturbation theory to include the intralayer interaction (recall that the interlayer
interaction is zero at long wavelength) corresponds to using the first Born approximation. This approximation traditionally
fails in the strong-coupling regime. However, the crossover physics of interest happens at low density
and we expect the first Born approximation to be reasonable. The perturbative parameter here is $Uk_Fd$ which 
must be small. This is true in our phase diagram for $U\lesssim 5$. However, 
the interlayer interaction that is responsible 
for the bound-state is taken into account essentially exactly through the integration over the 
bound-state wave function in Eq.~\eqref{veff}. We therefore expect our results to still be at least qualitatively 
correct for larger $U$ as well.

\subsection{Pauli Blocking Effect}
The above discussion was based on a binding energy for the dimer calculated from the pure two-body Schr{\"o}dinger
equation. The fact that the dimer must form in the proximity of other fermionic molecules was taken into account
by adding the Fermi energy and the intralayer repulsion. We now proceed by a different route in order to 
the background Fermi sea into account when forming the dimer. This can be seen as a 'top-down' approach, where the 
Fermi sea is considered a background on top of which a dimer forms. In contrast, a 'bottom-up' approach would have 
to consider the full $N$-body problem which is a much more difficult task. Our approach is the same as that used to
demonstrate the binding of Cooper pairs in the case of short-range interactions for two particles in the presence
of Fermi seas. Here it is known to be benificial to have the background, but this it is not a priory clear 
that the same holds with long-range dipolar interactions. Below, we calculate the energy in the presence of 
Fermi seas and show that the energy of the dimer goes up as the Fermi sea is increased. Thus taking the Pauli blocking effect into 
account to lowest order implies that the crossover could move in the phase diagram.
Notice that we have a Fermi sea background for {\it both} molecules that form the dimer. This is 
different from the polaron problem where only one of the particles has a background sea. Interetingly, the 
polaron problem in the bilayer setup with polar molecules has recently been studied by Klauwunn and Recati \cite{recati2011}.

Consider the momentum-space two-body Schr{\"o}dinger equation
\begin{align}
\frac{\hbar^2\bm k^2}{2\mu}\tilde\Psi(\bm k)+\int \frac{d^2\bm q}{(2\pi\hbar)^2} V_\perp(\bm k-\bm q)\tilde\Psi(\bm q)=E\tilde\Psi(\bm k),
\end{align}
where $\bm k$ is the relative momentum, which can be solved by discretization. We now include the effect of the background Fermi sea
on the dimer by blocking all states with momenta less than the Fermi wave vector $k_F$, i.e. we assume that $\tilde\Psi(\bm k)=0$ for $|\bm k|\leq k_F$.
We also assume that the dimer has zero center-of-mass momentum with respect to the Fermi sea, i.e. $\bm k_1=-\bm k_2$ so that the relative
momentum becomes the lab momentum. Here we are effectively assuming that the Fermi sea is inert in the sense that we neglect particle-hole pairs induced by the interactions. These correlations can be taken into account for instance through the variational ansatz employed for highly polarized 
Fermi gases with short-range interactions \cite{chevy2010}, suitably modified to the case of a balanced system which is what we are concerned with here. This will be explored in future work.

\begin{figure}[htb!]
\centering
  \epsfig{file=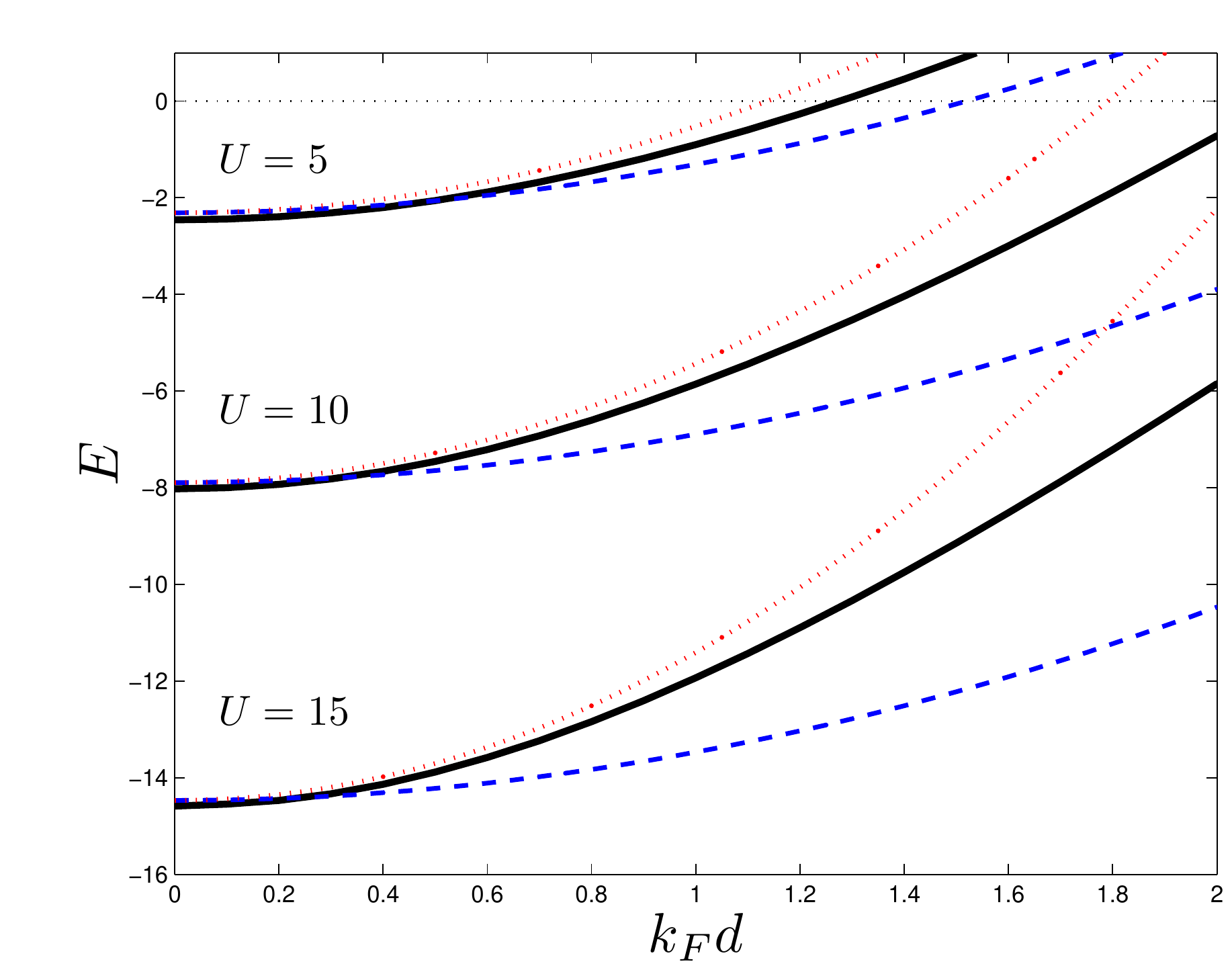,clip=true,scale=0.45}
  \caption{(color online) Dimer binding energy in units of $\hbar^2/Md^2$ for $U=5$, 10, and 15 as function of $k_Fd$. The solid (black) curve
  is the result of solving the momentum-space Schr{\"o}dinger equation with all states below $k_F$ blocked. The dashed (blue) curve is
  the binding energy obtained by using a variationally optimized Gaussian without blocking plus twice the Fermi energy (according to Eq.~\eqref{kine}), while the dotted (red) curve includes twice the Fermi energy and also the lowest-order correction to the potential energy at finite $k_F$, Eq.~\eqref{pote}.}
  \label{fig9}
\end{figure}

Before we discuss the numerical results, some limiting analytical expression can be obtained. Let us assume a Gaussian 
wave function which is a very good approximation for large $U$. Then the kinetic energy, $T$, in the Pauli blocked case 
is 
\begin{align}\label{kine}
\frac{\langle \Psi|T|\Psi\rangle}{\langle\Psi|\Psi\rangle}=\frac{\hbar^2}{2\mu b^2}\left[1+(k_Fb)^2\right],
\end{align}
where $b$ is the length scale of the Gaussian which can be obtained either variationally or within one of the approximation
schemes above. Here we have taken care to re-normalize the wave function when the $|\bm k|\leq k_F$ is cut.
This is an intuitively pleasing result, and we note that a similar expression can be obtained for the
exponential ansatz. 
The kinetic energy per fermionic molecule can then be written
\begin{align}
\frac{1}{2}\frac{\langle \Psi|T|\Psi\rangle}{\langle\Psi|\Psi\rangle}=\frac{\hbar^2}{4\mu b^2}+E_F.
\end{align}
This should be related to the estimation of the chemical potential through $\mu=E_F-\tfrac{1}{2}|E_0|$ as
shown on the right of Fig.~\ref{fig8}. This approximation corresponds to using Pauli blocking in the
kinetic energy while leaving the interaction energy term untouched. The matrix element of the potential in
momentum space for the Gaussian wave function is
\begin{align}
&\frac{\langle \Psi|V|\Psi\rangle}{\langle\Psi|\Psi\rangle}=4\pi b^2 e^{(k_Fb)^2}
\int_{k_F} \frac{d^2\bm k}{(2\pi)^2} e^{-b^2k^2}&\nonumber\\
&\int \frac{d^2\bm q}{(2\pi)^2}(-2\pi D^2)qe^{-qd-b^2q^2/2+b^2kq\cos\phi_q},&
\end{align}
where the factor $e^{(k_Fb)^2}$ comes from the re-normalization of the wave function, $\langle \Psi|\Psi\rangle=e^{-(k_Fb)^2}$. 
In the $k_F \to 0$ limit, the integral can be performed to yield the total energy given in Eq.~\eqref{hamil}. 
The corrections to the potential energy at finite $k_F$ can be obtained by expansion, which to lowest non-trivial order
yields 
\begin{align}\label{pote}
&\langle V \rangle =\langle V \rangle_{k_F=0}+\frac{1}{4}\frac{D^2}{d^3}(k_Fb)^2 \left(\frac{d}{b}\right)^5 g\left(\frac{b}{d}\right)&\\
&g(y)=-2 y+\left(1+y^2\right) \sqrt{2 \pi } \text{Erfc}\left[\frac{1}{\sqrt{2} y}\right]\times&\\\nonumber 
&\left(\text{Cosh}\left[\frac{1}{2 y^2}\right]+\text{Sinh}\left[\frac{1}{2 y^2}\right]\right).&
\end{align}
The function $g(y)$ is positive so that the energy increases with increasing $k_F$. 

In Fig.~\ref{fig9} we plot the exact solution including Pauli blocking (solid curve) for $U=5$, 10, and 15 as 
a function of $k_Fd$. As one 
would expect from the arguments above, the energy increases with $k_Fd$ in a quadratic fashion with a larger
slope for larger $U$. For comparison, we have also plotted the results of a Gaussian variational calculation
of the energy including the effect of blocking on the kinetic energy (dashed (blue) curve) and on 
both the kinetic and potential energy terms (dotted (red) curve). Recall again that the effect of blocking on the
kinetic energy is exact, whereas that on the potential part is only to lowest non-trivial order in $k_Fd$.
The dashed curve in Fig.~\ref{fig9}, which includes the blocking in the kinetic energy, 
overestimates the binding. The inclusion 
of blocking in the potential (dotted curve in Fig~\ref{fig9}) is seen to provide an improvement at small $k_Fd$ (particularly for larger $U$), 
but still underestimates the binding at large $k_Fd$. 
An expansion in large $k_Fd$ can also be done but this is not analytical
and provides only limited additional information. We will not pursue this further here.

\begin{figure}[htb!]
\centering
  \epsfig{file=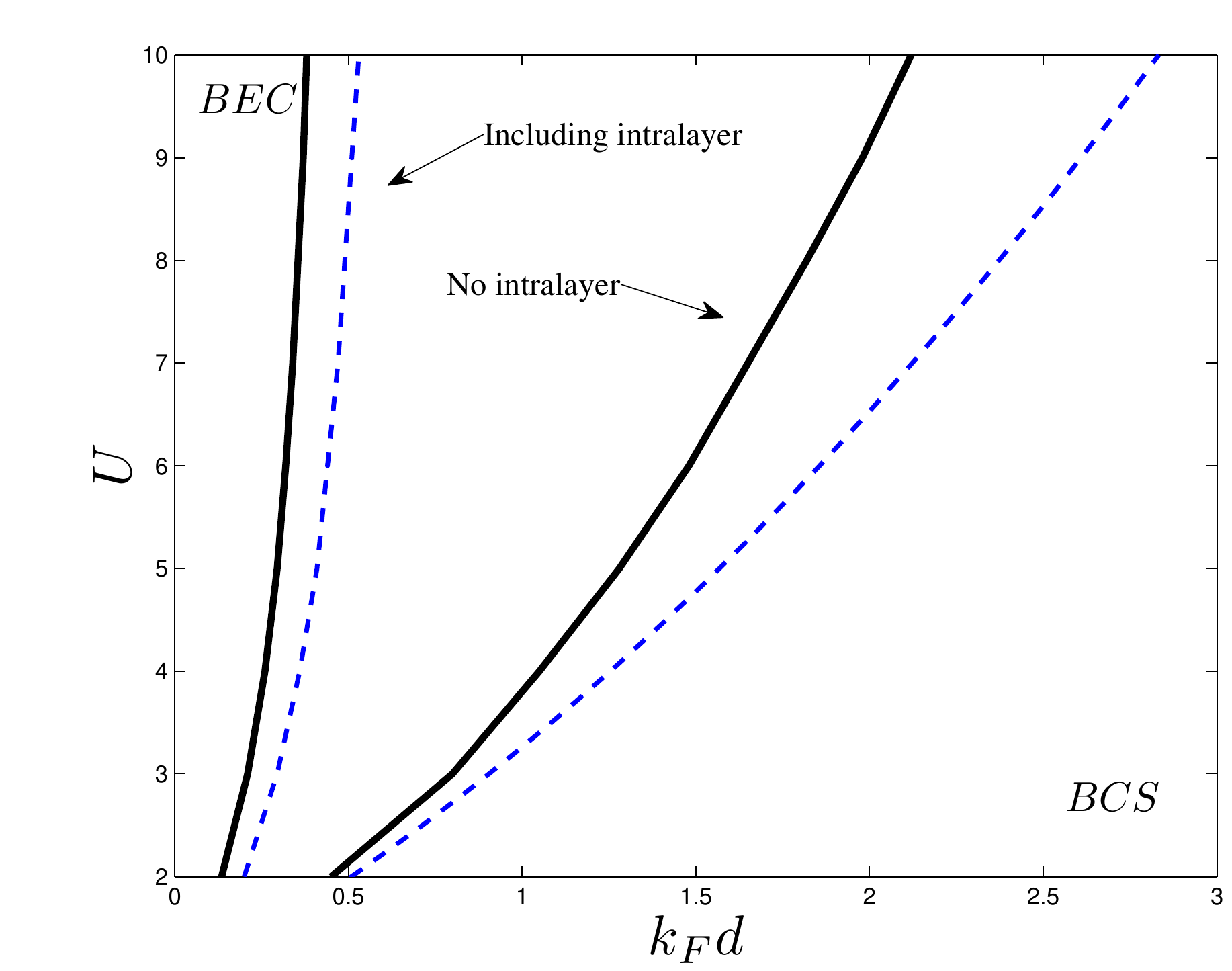,clip=true,scale=0.45}
  \caption{(color online) Phase diagram as in Fig.~\ref{fig8}. The solid (black) lines obtained from the momentum-space
  Schr{\"o}dinger equation including Pauli blocking of low momentum states with (right) and without (left) intralayer 
  interaction, whereas the dashed (blue) lines are obtained by adding the Fermi energy to the unblocked binding
  energy. The latter corresponds to the solid (blue) lines in Fig.~\ref{fig8}. Again we use $w/d=0.2$.}
  \label{fig10}
\end{figure}

The fact that decrease of the binding energy with increasing size of the background Fermi sea is faster than 
predicted by simple addition of the Fermi energy implies that the phase diagram in Fig.~\ref{fig9} must be 
suitably modified to account for the background effects.
Fig.~\ref{fig10} presents a comparison of the lines of vanishing chemical potential when including the Pauli blocking
through the exact solution of the momentum-space Schr{\"o}dinger equation, both with and without including the 
effects of the intralayer interaction in the long-wavelength limit through addition of $\tfrac{1}{2}nV_\text{eff}(0)$. 
This implies that the background environment reduces the importance of the dimer for the many-body 
physics.

\section{Conclusion and Outlook}
We have considered a bilayer system with fermionic polar molecules and in particular the
two-body bound dimer state that the system supports at any dipole strength. The energy and
wave function was calculated exactly and compared to various approximation schemes that 
provide convenient analytical expressions. We demonstrated that for dipole strength $U>1$, 
a suitably chosen Gaussian provide a very good analytical approximation in both coordinate- and
momentum-space. The effective dimer-dimer interaction was calculated within the different schemes and 
also found to be well approximated by using Gaussian two-body wave functions, which 
can be used, for instance, to study roton instabilities \cite{zinner2010}. In conclusion, 
we find that the Gaussian choice is both an accurate and a convenient one due to its simple
analytical properties.

The corresponding many-body physics of the bilayer was studied in the context of the 2D Fermi gas
where BCS-BEC crossover has been predicted for short-range interactions in the two-component system. 
This crossover depends sensitively on the two-body bound state. 
The bilayer system considered in the present paper is very similar since the layer index can be 
mapped to a spin quantum number. 
By determining where the 
chemical potential of the dipolar system vanishes, we could map out a mean-field phase diagram of 
the BCS and (quasi)-BEC regions. We find this diagram to be largely independent of our approximation 
schemes for the binding energy. Here we took the effects of the background Fermi sea on the 
dimer into account both through a simple addition of the Fermi energy, but also by using Pauli blocking
of all momenta below the Fermi momentum in the momentum-space Schr{\"o}dinger equation for the 
dimer binding energy. This showed that blocking causes a decrease in the binding with the size 
of the Fermi sea that is faster than naively expected by addition of the Fermi energy to the 
non-blocked two-body binding energy. 

In the bilayer with fermionic polar molecules there are both repulsive and
attractive parts. If we neglect the presence of the repulsive intralayer term in the bilayer, we 
recover the result from the short-range attractive 2D Fermi gas case. However, the long-range intralayer
repulsive drastically changes the picture and pushes the BCS-BEC crossover to much smaller
densities. This has clear analogs to electron-hole bilayers where the Coulomb interaction
provides the long-range part. Here we have included the effect of Pauli blocking on the 
two-body bound-state energy and demonstrated that this pushes the crossover to 
even smaller densities, implicating that the BCS paired state occupies the majority of 
the mean-field phase diagram.

To access the crossover region would require degenerate bilayer systems at very low densities. 
This is experimentally challenging since the critical temperature of this 2D setup is governed
by the BKT transition \cite{bkt1,bkt2}. The transition temperature is proportional to 
the superfluid density which can be depleted by both interaction and thermal effects. 
It is therefore driven to very small values at low density. 
In addition, we need dipole strengths 
in the range $U\sim 1-10$, which is higher than current experiments \cite{miranda2011}. 
Using molecules with a larger moment like LiCs or a different optical 
lattice setup is therefore necessary to probe the crossover.
Another possibility is for the system to 
develop a density wave \cite{sun2010,yamaguchi2010,zinner2011}. However, strongly bound 
dimers should change this picture and it is not clear exactly where the density wave transition 
sits. In any case, we expect it to occupy the large-density and large-strength part of the
phase diagram so the crossover physics should be accessible in the low-density and 
intermediate-strength region.

Two-body states are one important constituent in bilayer system. However, 

due to the long-range character of the dipolar interaction, we can have three- or more-body states
that are also bound, even though the particles are spatially separated and would
not exhibit bound states with zero- or short-range interaction. 
This is a very interesting feature as it is very different from the 
paradigmatic crossover BCS-BEC from weakly bound Cooper pairs to strongly bound bosonic two-body states 
which is driven by short-range two-body interactions. These questions will be 
pursued in future investigations.

\begin{acknowledgements}
NTZ gratefully acknowledges numerous discussions with D.-W. Wang and B. Wunsch. This work
was supported by the Danish Council for Independent Research $|$ Natural Sciences.
\end{acknowledgements}

\end{document}